%% file: main.tex
\documentclass[a4paper]{llncs}

\usepackage[latin1]{inputenc}
\usepackage[english]{babel}

\usepackage{graphicx}
\DeclareGraphicsExtensions{.jpg, .pdf, .ps}
\usepackage{xcolor}
\usepackage{subfig}
\usepackage{caption}

\usepackage{array}

\usepackage{amssymb}
\usepackage{amsmath}
\usepackage{enumitem}

\usepackage{multirow}

\usepackage{algorithm}
\usepackage{algpseudocode}
\algblockdefx[Product]{Product}{EndProduct}%
	[1]{\textsc{product}(#1)}{}
\algtext*{EndProduct}
\algblockdefx[Feedback]{Feedback}{EndFeedback}%
	[1]{\textsc{feedback}(#1)}{}
\algtext*{EndFeedback}
\algblockdefx[Serial]{Serial}{EndSerial}%
	[1]{\textsc{serial}(#1)}{}
\algtext*{EndSerial}

\def\techrep{}

\ifdefined\techrep
	\usepackage{fullpage}
\fi

\usepackage{url}
\usepackage[pdftex,bookmarks=true,colorlinks=true,linkcolor=blue,citecolor=blue,urlcolor=blue]{hyperref} 
\AtBeginDocument{}

\newcommand{\pop}{\mathop{\|}}
\newcommand{\sop}{\circ}

\newcommand{\bool}{\mathsf{Bool}}

\newcommand{\ski}{\mathsf{Id}}
\newcommand{\pred}{\mathsf{Pred}}
\newcommand{\fb}{\mathsf{feedback}}

\newcommand{\sel}{\mathsf{sel}}

\newcommand{\delete}[1]{}

\renewcommand{\implies}{\Rightarrow}

\title{Translating Hierarchical Block Diagrams into Composite Predicate Transformers\thanks{This work has been partially supported by the Academy of Finland, the U.S. National Science Foundation (awards \#1329759 and \#1139138), and by UC Berkeley's iCyPhy Research Center (supported by IBM and United Technologies).}}
\author{Iulia Dragomir\inst{1} \and Viorel Preoteasa\inst{1} \and Stavros Tripakis\inst{1,2}}

\institute{Aalto University, Finland \\
	\email{\{iulia.dragomir, viorel.preoteasa, stavros.tripakis\}@aalto.fi} \and
	University of California, Berkeley, USA}

\begin{document}

\maketitle

\begin{abstract}
\input{abstract.txt}
\end{abstract}

\input{intro}

\input{prelim}

\input{formalization}

\input{transfo}

\input{evaluation}

\input{rwork}

\input{conclusion}

\bibliographystyle{abbrv}
\bibliography{main}

\end{document}

%% file: abstract.txt
Simulink is the de facto industrial standard for designing embedded control
systems. When dealing with the formal verification of Simulink models, we face
the problem of translating the graphical language of Simulink, namely,
hierarchical block diagrams (HBDs), into a formalism suitable for
verification. In this paper, we study the translation of HBDs into the
compositional refinement calculus framework for reactive systems.
Specifically, we consider as target language an algebra of atomic predicate
transformers to capture basic Simulink blocks (both stateless and stateful),
composed in series, in parallel, and in feedback. For a given HBD, there are
many possible ways to translate it into a term in this algebra, with different
tradeoffs. We explore these tradeoffs, and present three translation
algorithms. We report on a prototype implementation of these algorithms in a
tool that translates Simulink models into algebra terms implemented in the
Isabelle theorem prover. We test our tool on several case studies
including a benchmark Simulink model by Toyota. We 
compare the three translation algorithms, with respect to size and readability of generated terms, simplifiability of the corresponding formulas, and other metrics.

%% file: intro.tex

\section{Introduction}
\label{sec:intro}


\delete{
Model-based design (MBD) is becoming widely used in the industry as an 
effective methodology to develop embedded systems. In MBD one starts
from a high-level model of the system (e.g., an automotive controller 
modeled as a set of differential equations), performs analysis on the model
(e.g., simulations) in order to find design defects, and applies successive
refinements to the model, all the way down to embedded code generation.
}


Simulink\footnote{\url{http://www.mathworks.com/products/simulink/}} is a widely used tool for modeling and simulating embedded control systems. 
Simulink uses a graphical language based on {\em hierarchical block diagrams}
(HBDs). HBDs are networks of interconnected {\em blocks}, which can be either
{\em basic} blocks from Simulink's libraries, or {\em composite} blocks 
({\em subsystems}), which are themselves HBDs. Hierarchy can be seen as the
primary {\em modularization} mechanism that Simulink offers, which allows to
master complexity of large models, improve their readability, and so on.

Our work seeks to develop methods and tools for {\em compositional} analysis
of HBDs, including but not limited to Simulink models. By ``compositional''
we mean exploiting the hierarchical structure of these diagrams, for instance,
reasoning about individual blocks and subsystems independently, and then
composing the results to reason about more complex systems. By ``analysis'',
we mean different types of checks, including exhaustive verification
(model-checking), but also more ``lightweight'' analyses such as
{\em compatibility checking}, which aims to check whether the connections
between two or more blocks in the diagram are valid, i.e., are the blocks
compatible.

We base our work on the {\em refinement calculus for reactive 
systems}~\cite{TripakisLHL11,PreoteasaT14}.
In this framework, systems are modeled as {\em predicate} and {\em property 
transformers}~\cite{back:wright:98,PreoteasaT14}. Open, non-deterministic, and non-input-receptive systems can be modeled in the framework. 
{\em Serial}, {\em parallel}, and {\em feedback} composition operators 
can be used
to form more complex systems from simpler ones. Compatibility can be
checked during such compositions. Both safety and liveness properties can
be expressed in the framework (e.g., using LTL) and used for verification. 
In addition, the framework includes the notion of {\em refinement}, a binary
relation between components, which characterizes {\em substitutability}
(when can a component replace another one while preserving system properties).
Refinement has multiple usages, including compositional and incremental design,
and reusability. This makes the framework compelling for application on tools
like Simulink, which have a naturally compositional hierarchical language.

In order to use refinement calculus\footnote{but also any other framework
that relies on an algebra of atomic components composed with operators such
as serial, parallel, and feedback} with Simulink, we need to translate the
graphical language of Simulink (HBDs) into the algebra of composite 
transformers. This translation raises interesting problems, and these are
the topic of this paper.

\begin{figure}[t] 
	\centering 
	\subfloat[$\fb_{a}(P_A \sop (P_B \pop \ski))$]{\label{fig:mex_a}
		\parbox{3.8cm}{\centering\includegraphics[scale=0.82]{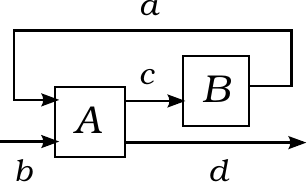}}} \quad
	\subfloat[$\fb_{c}((P_B \pop \ski) \sop P_A)$]{\label{fig:mex_b}
		\parbox{3.8cm}{\centering\includegraphics[scale=0.82]{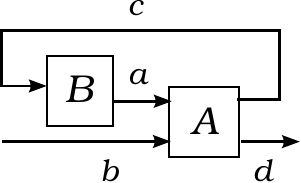}}} \quad
	\subfloat[$\fb_{a,c}(P_A \pop P_B)$]{\label{fig:mex_c}
		\raisebox{0.18cm}{\parbox{3.5cm}{\centering\includegraphics[scale=0.82]{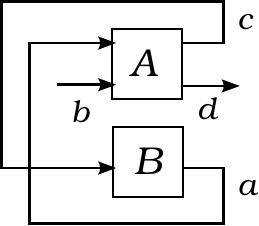}}}}
	\caption{Three ways to view and translate the same block diagram.}\label{fig:mex}
\end{figure}

To illustrate some of the questions that arise, consider the block diagram in Fig.~\ref{fig:mex_a}.
Let $P_A$ and $P_B$ be transformers modeling the blocks $A$ and $B$ 
in the diagram. How should we compose $P_A$ and $P_B$ in order to get a 
transformer that represents the entire diagram?
As it turns out, there are many such compositions possible.
One option is to compose first $P_A$ and $P_B$ in series, and then compose
the result in feedback, following Fig.~\ref{fig:mex_a}.
This results in the composite transformer $\fb_{a}(P_A \sop (P_B \pop \ski))$,
where $\sop$ is composition in series, $\pop$ in parallel, and
$\fb_x$ is feedback applied on port $x$. $\ski$ is the transformer
representing the identity function. $A$ has two outputs and $B$ only one
input, therefore to connect them in series we first form the parallel
composition $P_B\pop\ski$, which represents a system with two inputs.

Another option is to compose the blocks in series in the opposite order,
$P_B$ followed by $P_A$, and then apply feedback. This results in the
transformer $\fb_{c}((P_B \pop \ski) \sop P_A)$. A third option
is to compose the two blocks first in parallel, and then apply feedback
on the two ports $a,c$. This results in the transformer
$\fb_{a,c}(P_A \pop P_B)$.
Although semantically equivalent, these three transformers have
different computational properties.

Clearly, for complex diagrams, there are many possible translation
options. In this paper we study these options in depth. Our main contributions
are the following.
First, we present three different translation strategies: {\em feedback-parallel}
translation which forms the parallel composition of all blocks, and then
applies feedback; {\em incremental} translation which orders blocks 
topologically and composes them one by one; and {\em feedbackless} translation,
which avoids feedback composition altogether, provided the original block
diagram has no algebraic loops.
Second, we discuss the tradeoffs of these strategies, in terms of several
metrics, including the size of the resulting expressions, readability, and
also {\em simplifiability}, which can generally be defined as how 
easy/automatic it is to simplify the expressions into simple formulas 
characterizing the overall system. Third, we report on a prototype tool
which implements the three translation strategies. The tool takes as
input hierarchical Simulink models and generates composite predicate
transformers implemented in the Isabelle proof assistant\footnote{\url{https://isabelle.in.tum.de/}}.
Isabelle or other tools are then used to simplify the corresponding formulas,
to perform verification, etc.
We evaluate and compare the three translation strategies on several case 
studies, including a Fuel Control System benchmark by Toyota~\cite{jin2014ARCH,JinDKUB14}.

\noindent\paragraph{Paper structure.} In \S\ref{sec:HBD} we present the basic modeling notions in Simulink for designing hierarchical systems and their semantics. We illustrate these notions on a toy running example, a 1-step counter. \S\ref{sec:rcrs} depicts the predicate transformer algebra containing atomic and composite terms and three composition operators. Some examples on how to formalize atomic Simulink blocks with predicate transformers are presented. In \S\ref{sec:transformation}, the three translation strategies of Simulink hierarchical block diagrams to composite predicate transformers are described and illustrated on examples. We report on the tool support in \S\ref{sec:impl_eval} and we validate the approach on two case studies. We discuss the benefits and drawbacks of each translation strategy with respect to different metrics. Finally, we discuss the related work before concluding. 

%% file: prelim.tex

\section{Hierarchical Block Diagrams in Simulink}
\label{sec:HBD}

A Simulink HBD is a network of blocks interconnected by wires. Blocks can be either {\em basic} blocks from the Simulink libraries, or {\em composite} blocks ({\em subsystems}). A basic block is described by: (1) a label, (2) a list of parameters, (3) a list of in- and out-ports,
(4) a vector of state variables with predefined initial values (i.e., the local memory of a block) and (5) functions to compute the outputs and next state variables. The outputs are computed from the inputs, current state and parameters. State variables are updated by a function with the same arguments. Subsystems are defined by their label, list of in- and out-ports, and the list of block instances that they contain -- both atomic and composite.

\delete{
Simulink defines a data flow computational model, where signals (i.e. data) are transferred through wires from one block to another. The data is transferred from the outport of a block to the connected input of another one. It is possible for the outport of one block to be connected to several inputs of different blocks by split wires. Therefore, Simulink describes a multicast communication pattern: the same signal is transferred to all modeled receptors in one step. 
}

Simulink allows to model both discrete and continuous-time blocks. For example, \textsf{UnitDelay} is a discrete-time block which outputs at step $n$ the input at step $n-1$. An \textsf{Integrator} is a continuous-time block whose output is described by a differential equation solved with numerical methods. We interpret a Simulink model as a discrete-time model (essentially an input-output state machine, possibly infinite-state) which evolves in a sequence of discrete steps. Each step has duration $\Delta t$, which is a parameter (user-defined or automatically computed by Simulink based on the blocks' time rates). In this paper we consider single-rate models. 
We also assume that Simulink models are free from {\em algebraic loops}, that
is, feedback loops resulting in instantaneous cyclic dependencies. We can have feedback loops, but they must be ``broken'' by blocks such as \textsf{UnitDelay}
or \textsf{Integrator}, as in the example that follows.

\begin{figure}[!t]
	\centering
	\subfloat[Hierarchical Block Diagram]{\label{fig:counter_hbd}
		\includegraphics[scale=0.70]{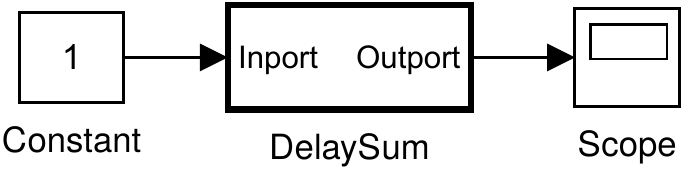}} \qquad
	\subfloat[\textsf{DelaySum} Subsystem]{\label{fig:counter_bd}
		\includegraphics[scale=0.70]{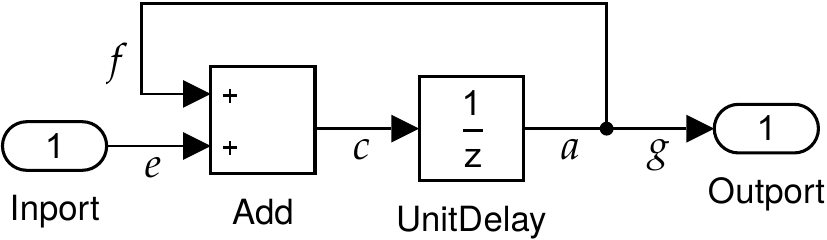}}
	\caption{Simulink model of a counter with step 1.}\label{fig:counter}
\end{figure}

\paragraph{Running example.} Throughout the paper we illustrate our methods 
using a simple example of a counter, shown in Fig.~\ref{fig:counter}. This is 
a hierarchical (two-level) Simulink model. The top-level diagram (Fig.~\ref{fig:counter_hbd}) contains three block instances: the step of the counter as a \textsf{Constant} basic block, the subsystem \textsf{DelaySum}, and the \textsf{Scope} basic block which allows to view simulation results.
The subsystem \textsf{DelaySum} (Fig.~\ref{fig:counter_bd})
contains a \textsf{UnitDelay} block instance which represents the state
of the counter. \textsf{UnitDelay} can be specified by the formula $a = s\land s' = c$, where $c$ is the input, $a$ the output, $s$ the current state and $s'$ the next state variable. We assume that $s$ is initially $0$. 
The $\mathsf{Add}$ block instance adds the two input values and outputs the result in the same time step: $c=f+e$. The {\em junction} after
 wire $a$ (black dot in the figure) can be seen as a basic block duplicating
(or {\em splitting}) its input to its two outputs: $f=a\land g=a$.


%% file: formalization.tex
\section{Atomic and Composite Predicate Transformers}
\label{sec:rcrs}

Monotonic predicate transformers \cite{dijkstra:75} (MPTs) are an expressive formalism, used within the context of programming languages to model non-determinism, correctness (both functional correctness and termination), and refinement \cite{back:wright:98}.
In this section we show how MPTs can also model input-output systems such as state machines, and by extension also capture the semantics of block diagrams {\em \`a la} Simulink. The idea is to represent the computation step of a state machine using an MPT which has the inputs of the state machine and the {\em current state} variable as inputs, and the outputs of the state machine and the {\em next state} variable as outputs.
We also discuss MPT composition operators.

\subsection{Monotonic Predicate Transformers for Modeling Systems}
\label{subsec:pt}

A {\em predicate} on $\Sigma$ is a function $q:\Sigma\to\bool$. Predicate $q$ can also be seen as a subset of $\Sigma$: for $\sigma\in\Sigma$, $\sigma$ belongs to the subset iff $q(\sigma)$ is true. 
\delete{(we use $.$ for function application and write $q.\sigma$ instead of $q(\sigma)$).}
Predicates can be ordered by the subset relation: we write $q\le q'$ if predicate $q$, viewed as a set, is a subset of $q'$.
$\pred(\Sigma)$ denotes the set of predicates $\Sigma\to\bool$. 

A {\em predicate transformer} is a function $S: (\Sigma'\to\bool)\to(\Sigma\to\bool)$, or equivalently, $S: \pred(\Sigma')\to\pred(\Sigma)$. $S$ takes a predicate on $\Sigma'$ and returns a predicate on $\Sigma$. $S$ is {\em monotonic} if $\forall q,q' : q\le q' \implies S(q) \le S(q')$.

Traditionally, MPTs have been used to model sequential programs using weakest precondition semantics. Given a MPT $S: (\Sigma'\to\bool)\to(\Sigma\to\bool)$, and a predicate $q':\Sigma'\to\bool$ capturing a set of {\em final} states, $S(q')$ captures the set of all {\em initial} states, such that if the program is started in any state in $S(q')$, it is guaranteed to finish in some state in $q'$. But this is not the only possible interpretation of $S$. $S$ can also model input-output systems. For instance, $S$ can model a {\em stateless} system with a single inport ranging over $\Sigma$, and a single outport ranging over $\Sigma'$. Given a predicate $q'$ characterizing a set of possible {\em output values}, $S(q')$ characterizes the set of all {\em input values} which, when fed into the system, result in the system outputting a value in $q'$. As an example, the identity function can be modeled by the MPT
$\ski:\pred(\Sigma)\to\pred(\Sigma)$, defined by $\ski(q)= q$, for any $q$.

MPTs can also model {\em stateful} systems. For instance, consider the $\mathsf{UnitDelay}$ described in \S\ref{sec:HBD}. Let the input, output, and state variable of this system range over some domain $\Sigma$. Then, this system can be modeled as a MPT $S: \pred(\Sigma\times\Sigma)\to\pred(\Sigma\times\Sigma)$. The Cartesian product $\Sigma\times\Sigma$ captures pairs of {\em (input, current state)} or {\em (output, next state)} values. Intuitively, we can think of this system as a function which takes as input $(x,s)$, the input $x$ and the current state $s$, and returns $(y,s')$, the output and the next state $s'$, such that $y=s$ and $s'=x$. The MPT $S$ can then be defined as follows:
$$
S(q) = \{ (x,s) \mid (s,x)\in q \}.
$$
In the definition above we view predicates $q$ and $S(q)$ as sets.

Syntactically, a convenient way to specify systems is using formulas on input, output, and state variables. For example, the identity system can be specified by the formula $y=x$, where $y$ is the output variable and $x$ is the input. The \textsf{UnitDelay} system can be specified by the formula $y=s\land s'=x$. We next introduce operators which define MPTs from predicates and relations.

For a predicate $p:\Sigma\to\bool$ and a relation $r:\Sigma\to\Sigma'\to\bool$, we define the {\em assert} MPT, $\{p\}:\pred(\Sigma)\to\pred(\Sigma)$, and the {\em non-deterministic update} MPT, $[r]:\pred(\Sigma')\to\pred(\Sigma)$, where:
$$
\begin{array}{lll}
\{p\}(q) = (p\land q) & \mbox{ and } &
[r](q) = \{\sigma \ | \ \forall \sigma' : \sigma' \in r(\sigma) \Rightarrow \sigma' \in q\} 
\end{array}
$$

Transformer $\{p\}$ is used to model non-input-receptive systems, that is, systems where some inputs are illegal~\cite{TripakisLHL11}.
$\{p\}$ constrains the inputs so that they must satisfy predicate $p$. 
It accepts only those inputs and behaves like the identity function. That is, $\{p\}$ models a partial identity function, restricted to the domain $p$.
Transformer $[r]$ models an input-receptive but possibly non-deterministic system.
Given input $\sigma$, the system chooses non-deterministically some output $\sigma'$ such that $\sigma' \in r(\sigma)$ is true. If no such $\sigma'$ exists, then the system behaves {\em miraculously} \cite{back:wright:98}. In our framework we ensure non-miraculous behavior, as explained below.

To model basic Simulink blocks, we often combine $\{p\}$ and $[r]$ using the {\em serial composition} operator $\sop$, which for predicate transformers is simply function composition. Given two MPTs $S:\pred(\Sigma_2)\to\pred(\Sigma_1)$ and $T:\pred(\Sigma_3)\to\pred(\Sigma_2)$, their serial composition
$(S\sop T):\pred(\Sigma_3)\to\pred(\Sigma_1)$ is defined as $(S\sop T)(q) = S(T(q))$.

For example, consider a system with two inputs $x,y$ and one output $z$, performing the division $z = \frac{x}{y}$. We want to state that division by zero is illegal, and therefore, the system should reject any input where $y=0$. This system can be specified as the MPT
$$
\mathsf{Div} = \{\lambda x,y: y \not= 0\}\circ[\lambda (x,y),z : z = \frac{x}{y}]
$$
where we employ lambda-notation for functions. Note that $[\lambda (x,y),z : z = \frac{x}{y}]$ alone is not enough to capture $\mathsf{Div}$, because it allows illegal inputs where $y=0$ (and then behaves miraculously). The same is true for $[\lambda (x,y), z : y \not= 0 \land z = \frac{x}{y}]$. To ensure non-miraculous behavior, we always model non-input-receptive systems using a suitable
assert transformer $\{p\}$.

For a function $f:\Sigma\to\Sigma'$ the {\em functional update} $[f]:\pred(\Sigma')\to\pred(\Sigma)$ is defined as $[\lambda \sigma,\sigma':\sigma'=f(\sigma)]$ and we have 
$$
[f](q) = \{\sigma \ | \  f(\sigma) \in q \} = f^{-1}(q) 
$$

{\em Functional predicate transformers} are of the form $\{p\}\circ[f]$, and {\em relational predicate transformers} are of the form $\{p\}\circ[r]$, where $p$ is a predicate, $f$ is a function, and $r$ is a relation. {\em Atomic predicate transformers} are either functional or relational transformers.
$\mathsf{Div}$ is a functional predicate transformer which can also be written as
$
\mathsf{Div} = \{\lambda x,y: y \not= 0\}\circ[\lambda x,y : \frac{x}{y}].
$



For assert and update transformers based on Boolean expressions we introduce a simplified notation that avoids lambda abstractions. If $P$ is Boolean expression on some variables $x_1,\ldots,x_n$, then $\{x_1,\ldots,x_n:P\}$ denotes the assert transformer $\{\lambda x_1,\ldots,x_n:P\}$. Similarly if $R$ is a Boolean expression on variables $x_1,\ldots,x_n,y_1,\ldots,y_k$ and $F$ is a tuple of numerical expressions on variables $x_1,\ldots,x_n$, then $[x_1,\ldots,x_n\leadsto y_1,\ldots,y_k:R]$ and $[x_1,\ldots,x_n\leadsto F]$ are notations for $[\lambda (x_1,\ldots,x_n), (y_1,\ldots,y_k):R]$ and $[\lambda x_1,\ldots,x_n : F]$, respectively. With these notations the $\mathsf{Div}$ transformer becomes:
$$
\mathsf{Div} = \{x,y: y \not= 0\}\circ[x,y \leadsto \frac{x}{y}]
$$

\subsection{More Examples of Atomic Predicate Transformers}

We already saw how to model basic Simulink blocks like $\ski$ and $\mathsf{Div}$. Next we provide more examples, such as constants, delays, and integrators. A constant block parameterized by constant $c$ has no input, and a single output equal to $c$. As a predicate transformer the constant block has as input the
empty tuple $()$, and outputs the constant $c$:
$$
\mathsf{Const}(c) = [()\leadsto c]
$$
 
The unit delay block is modeled as the atomic predicate transformer 
$$
\mathsf{UnitDelay} = [x,s\leadsto s,x]
$$

Simulink includes continuous-time blocks such as the {\em integrator}, which computes the integral $\int _0^x f$ of a function $f$. 
Simulink uses different integration methods to simulate this block.
We use the Euler method with fixed time interval $\mathit{dt}$, a parameter. If $x$ is the input, $y$ the output, and $s$ the state variable of the integrator, then $y=s$ and $s'=s+x\cdot\mathit{dt}$. Therefore, the integrator can be modeled as the MPT
$$
\mathsf{Integrator}(\mathit{dt}) = [x,s\leadsto s, s + x\cdot \mathit{dt}]
$$

All other Simulink atomic blocks fall within these cases discussed above. Relation (\ref{def:blocks}) introduces the definitions of some blocks that we use in our examples.

\begin{equation}\label{def:blocks}
\mathsf{Add}  = [x, y \leadsto x + y ] \ \ \ \ \ 
\mathsf{Split}  =  [x\leadsto x,x]  \ \ \ \ \ 
\mathsf{Scope} =   \ski  
\end{equation}

\subsection{Composite Predicate Transformers}
\label{subsec:pte}

Basic Simulink blocks are modeled using atomic predicate transformers. To model arbitrary block diagrams, we use {\em composite predicate transformers} (CPTs). These are expressions over the atomic predicate transformers using serial, {\em parallel}, and {\em feedback} composition operators.
Serial composition $\sop$ was already introduced in \S\ref{subsec:pt}. \ifdefined\techrep Next we define the parallel and the feedback operators. \else Next we briefly introduce parallel and feedback composition, focusing only on how we use them the context of this paper. The complete formal definitions of these operators can be found in \cite{report} and in the Isabelle theories\footnote{\url{http://users.ics.aalto.fi/iulia/sim2isa.shtml}} that accompany this submission.\fi

\ifdefined\techrep
For $S:\pred(Y)\to\pred(X)$, and $T:\pred(Y')\to \pred(X')$, the parallel
composition of $S$ and $T$, denoted $S \pop T:\pred(Y\times Y')\to \pred(X\times X')$ is defined as
$$
  (S \pop T)(q) = \{ (x,x') \ | \ \exists p, p': p\times p' \le q \land x \in S(p) \land x' \in T(p') \}
$$
\else 
For two MPTs $S:\pred(Y)\to\pred(X)$ 
and $T:\pred(Y')\to \pred(X')$,
their parallel composition is the MPT $S \pop T:\pred(Y\times Y')\to \pred(X\times X')$. 
If $S =\{p\} \sop [f] $ and $T = \{p'\} \sop [f']$ are functional predicate transformers, then 
\begin{equation}\label{eq:parallel}
	S \pop T = \{x,x':p(x) \land p(x')\} \sop [x,x'\leadsto f(x), f'(x')]
\end{equation}
(\ref{eq:parallel}) states that input $(x,x')$ is legal for $S\pop T$ if $x$ is a legal input for $S$ and $x'$ is a legal input for $T$, and that the output of $S \pop T$ is the pair $(f(x),f'(x'))$.
\fi

\ifdefined\techrep
For $S:\pred(U\times Y) \to \pred(U\times X)$, the feedback of $S$, denoted 
$\fb(S):\pred(Y)\to\pred(X)$ is defined as
$$
\fb(S) = [x\leadsto x,x]\sop 
      (\sel(S) \pop \ski) \sop S \sop [v,y\leadsto y]  
$$
where
$$
\sel(S) = \{x:\exists u : (u,x) \in S(\top)\} \sop 
  [x\leadsto u, x: (u,x) \in S(\top) ]\sop S \sop [v,y \leadsto v]
$$
In this definition $\top$ is the total predicate ($\top.x$ is true for all $x$), and $S(\top)$ is the set of all legal inputs of $S$. 
The feedback operation is designed such that it provides the expected result
for a PT $S$ when its first output $v$ depends only on the
second input $x$. Intuitively this case is represented in Fig.\ref{fig:fb_c}. If $S$ has this special form, then $\mathsf{sel}(S)$ selects the component $T'$ ($\mathsf{sel}(S) = T'$). The component $T$ is $S \circ [v,y\leadsto y]$, and the feedback of $S$ is
represented in Figure~\ref{fig:fb_d}: 
$[x\leadsto x,x] \sop (T' \pop \ski) \circ T$.
Proper Simulink diagrams (those without algebraic loops) satisfy this property.

\begin{figure}[!t]
	\centering
	\subfloat[]{\label{fig:fb_a}
		\includegraphics{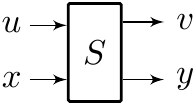}} \quad	
	\subfloat[]{\label{fig:fb_b}
		\includegraphics{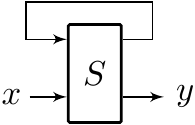}} \quad	
	\subfloat[]{\label{fig:fb_c}
		\includegraphics{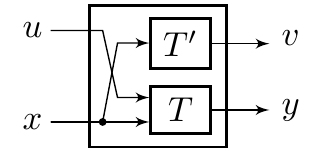}} \quad	
	\subfloat[]{\label{fig:fb_d}
		\includegraphics{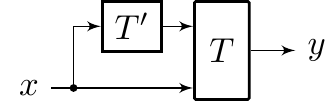}}
	\caption{(a) MPT $S$, (b) $\fb(S)$, (c) special case of $S$, (d) feedback of (c).}
	\label{fig:feedback}
\end{figure}
\else
For $S:\pred(U\times Y) \to \pred(U\times X)$ as in Fig.~\ref{fig:fb_a}, the feedback of $S$, denoted $\fb(S):\pred(Y)\to\pred(X)$ is obtained by connecting output $v$ to input $u$ (Fig.~\ref{fig:fb_b}). The feedback operation is designed such that it provides the expected result for a component $S$ when its first output $v$ depends only on the second input $x$ (Fig.~\ref{fig:fb_c}). Proper Simulink diagrams (those without algebraic loops) satisfy this property. The result of the feedback in this case is depicted in Fig.~\ref{fig:fb_d}.

\begin{figure}[!t]
	\centering
	\subfloat[]{\label{fig:fb_a}
		\includegraphics{figures/feedback-a}} \quad	
	\subfloat[]{\label{fig:fb_b}
		\includegraphics{figures/feedback-b}} \quad	
	\subfloat[]{\label{fig:fb_c}
		\includegraphics{figures/feedback-c}} \quad	
	\subfloat[]{\label{fig:fb_d}
		\includegraphics{figures/feedback-d}}
	\caption{(a) MPT $S$, (b) $\fb(S)$, (c) special case of $S$, (d) feedback of (c).}
	\label{fig:feedback}
\end{figure}

If $S$ is a functional predicate transformer which has the form of Fig.~\ref{fig:fb_c}, i.e., $S = \{p\} \sop [u,x \leadsto f'(x),f(u,x)]$, then its feedback satisfies:
\begin{equation}\label{eq:feedback}
  	\fb (S)  = 
  	   \{x: p(f'(x),x)) \} \sop [x\leadsto f(f'(x),x)]
\end{equation}
Input $x$ is legal for the feedback if $p(f'(x),x)$ is true, and the output for $x$ is $f(f'(x), x)$.
Unfortunately, it is non-trivial to automatically identify when does an
arbitrary transformer $S$ have the required form, i.e., to identify $f$ and
$f'$ above. For that reason, our implementation uses a more general formula
for feedback, which contains quantifiers
and generally results in relational, rather than functional,
transformers~\cite{report}.

For serial composition of two functional predicate transformers $S = \{p\}\sop [f]$ and $T = \{p'\}\sop[f']$  we have the following property:
\begin{equation}\label{eq:serial}
  S \sop T = \{p \land (p' \sop f)\} \sop   [f'\sop f]
\end{equation}
In (\ref{eq:serial}) input $x$ is legal if $x$ is legal for $S$ and the output of $S$, $f(x)$, is legal for $T$, i.e., $(p\land (p' \circ f))(x) = p(x) \land p'(f(x))$ is true. The output of $S \sop T$ is $(f' \circ f)(x) = f'(f(x))$.
\fi

\ifdefined\techrep
Next we introduce two lemmas for unfolding the composition operators when they are applied to relational and functional predicate transformers.
For serial and parallel compositions we have:
\begin{lemma} \label{lem:serial-prod}
For $p$, $p'$, $r$, $r'$, $f$, and $f'$ of appropriate types we have:
\begin{enumerate}[label=(\roman*)]
\item\label{lem:sp_a} $\{p\}\circ [r]\circ \{p'\}\circ[r'] = \{ x: x \in p \land (\forall y: y \in r(x)\Rightarrow y \in p')\} \circ [r\circ r']$
\item\label{lem:sp_b} $\{p\}\circ [f]\circ \{p'\}\circ[f'] = \{p \land (p' \circ f)\} \circ [f'\circ f]$
\item\label{lem:sp_c} 
$(\{p\}\circ[r]) \pop (\{p'\}\circ [r']) = \{x,y : x \in p \land y \in p'\}\circ
[x,y\leadsto u,v: u \in r(x) \land v \in r'(y)]$
\item\label{lem:sp_d}
$(\{p\}\circ[f]) \pop (\{p'\}\circ [f']) = \{x,y: x \in p \land y\in p'\}\circ
[x,y\leadsto f(x),f'(y)]$
\end{enumerate}
\end{lemma}

For feedback we have:
\begin{lemma} \label{lem:feedback} For $p$, $r$, $f$ and $f'$ of appropriate types we have:
\begin{enumerate}[label=(\roman*)]
\item\label{lem:fb_a} $\fb (\{p\} \circ [r]) = $ \\
$
  \begin{array}{l}
    \quad \{ x : (\exists u : (u, x) \in p) \land
     (\forall a : (\exists u : (u, x) \in p\land (\exists y : (a, y) \in r(u, x)))
     \Rightarrow (a, x) \in p) \} \\ 
     \quad \circ\ 
    [x \leadsto y : (\exists v : (\exists u : (u, x) \in p \land (\exists y:
    (v, y) \in r(u, x))) \land  (v, y) \in r(v, x))]
  \end{array}
$
\item\label{lem:fb_b}
$
  	\fb (\{p\} \sop 
  	   [u,x \leadsto f'(x),f(u,x)])  = 
  	   \{x: (f'(x),x) \in p) \} \sop [x\leadsto f(f'(x),x)]
$
  \end{enumerate}
\end{lemma}

Lemma \ref{lem:feedback}.\ref{lem:fb_a} gives a general formula for unfolding the feedback of a relational predicate transformer, and Lemma \ref{lem:feedback}.\ref{lem:fb_b} shows that the feedback works as expected when applied to a functional predicate transformer as in Fig.\ref{fig:fb_c}.
Although feedback operations applied to a proper Simulink diagrams are 
always of the form from Fig.\ref{fig:fb_c}, when translating arbitrary diagrams  {\em compositionally} to CPTs, we may not be able to identify the components $f'$ and $f$. Because of this, to simplify the CPTs, we need to use Lemma \ref{lem:feedback}.\ref{lem:fb_a}, and rely on a powerful simplification mechanism that will eliminate the quantifiers. Another problem with 
Lemma \ref{lem:feedback}.\ref{lem:fb_a} is that, although atomic Simulink blocks are
functional, after applying the feedback we obtain a relational PT.
\fi

As an illustration of how CPTs are used in this work, 
consider our running example (Fig.~\ref{fig:counter}).
An example translation of the \textsf{DelaySum} subsystem and of
the top-level Simulink model yields the following two CPTs:
\begin{equation}\label{eq:counter:def}
\begin{array}{lll}
\mathsf{DelaySum} & = &
  \fb((\mathsf{Add}\pop\ski)\sop \mathsf{UnitDelay}\sop(\mathsf{Split} \pop \ski))\\[0.7ex]
\mathsf{Counter} &=& (\mathsf{Const}(1) \pop \ski)\sop \mathsf{DelaySum} \sop (\mathsf{Scope} \pop \ski)
\end{array}
\end{equation}
The $\ski$ transformers in these definitions are for propagating the state introduced by the unit delay. 
\ifdefined\techrep
Expanding the definitions of the atomic blocks, and applying Lemmas \ref{lem:serial-prod}.\ref{lem:sp_d}, \ref{lem:feedback}.\ref{lem:fb_b}, and \ref{lem:serial-prod}.\ref{lem:sp_b} we obtain the {\em simplified} definitions:
\else
Expanding the definitions of the atomic blocks, and applying properties (\ref{eq:parallel}), (\ref{eq:feedback}), and (\ref{eq:serial}) we obtain the {\em simplified} definitions:
\fi
\begin{equation}\label{eq:counter:simp}
\begin{array}{lll}
\mathsf{DelaySum}  =  [x,s\leadsto s, s + x]
& \qquad \mbox{and} & \qquad
\mathsf{Counter} = [s\leadsto s, s + 1]
\end{array}
\end{equation}

Our goal is to perform automatically, (a) the translation from Simulink
diagrams (Fig.~\ref{fig:counter}) to CPTs (Defs.(\ref{eq:counter:def}));
and (b) the expansion and simplification from (\ref{eq:counter:def}) to
(\ref{eq:counter:simp}).
In the next section we discuss problem (a). For (b), we use Isabelle
and related tools.

%% file: transfo.tex

\section{From Hierarchical Block Diagrams to Composite Predicate Transformers}
\label{sec:transformation}

We want to translate HBDs to CPTs.\ifdefined\techrep\footnote{
We mention that this model-to-text translation is generic enough to accept as input any graphical hierarchical block diagrams, and it is not restricted to Simulink.} \else\  \fi 
 As illustrated in the introduction, this mapping is not unique: for a given HBD, there are many possible CPTs that we could generate. As we shall see in \S\ref{sec:impl_eval}, these CPTs, although semantically equivalent, have different simplifiability properties.
In this section, we describe three different translation strategies.

\ifdefined\techrep
The overall translation flow consists in the following steps:
\begin{enumerate}
	\item Parsing of the Simulink model diagram(s) in order to extract the relevant information about the system structure (i.e. blocks, connectors) and its processing (e.g., wire instantiation).
	\item Mapping of each elementary atomic block to an atomic property transformer. These PTs are generated at runtime based on the customizable number of inputs and outputs of atomic blocks. Examples of PTs for atomic blocks have been provided in \S\ref{subsec:pt}.
	\item Building CPTs from hierarchical structures by applying different translation strategies on components' compositions.
	\item Producing a complete Isabelle theory subjected to analysis and verification.
\end{enumerate}
This process is automatized by a compiler which is described in \S\ref{subsec:tool}. In the following we focus on the translation strategies for HBDs.
\fi

\ifdefined\techrep
\subsection{Internal Representation of HBDs}
\label{subsec:repr}

In order to manipulate Simulink HBDs, we define an internal representation of a Simulink model as a data structure in our tool. This internal representation is faithful to the graphical one. Each block instance $A$ consists of a list of inputs $A.in$, a list of outputs $A.out$ and a predicate transformer $A.cpt$. Moreover, if the block instance corresponds to a subsystem it will also enclose the list of contained components. By \textit{component} we refer in the following to block instances as described by the internal representation. 

The lists of inputs/outputs are made of variables (i.e. names), which in case of vector data are indexed over the number of elements in the vector. Wires between block instances are modeled by variables name matching, i.e., the output of a component has the same name as the input of its target. We use the following operations on input/output lists: 
(1) $+$ is list concatenation, 
(2) $\cap$ is list intersection which preserves the order of elements from the first operand, and
(3) $\setminus$ is list difference which again preserves the order of elements of the first operand.
The CPT for a component either is of a predefined type in case of an atomic block or is obtained by applying composition operators on the contained components' CPTs. 
\fi

\delete{
We mention that before computing a subsystem's CPT two algorithms are executed on its diagram in order to optimize the result. Firstly, we identify and build disconnected islands of components which are put in parallel for obtaining the diagram's CPT. Secondly, the components of each island are topologically sorted based on their dependencies. 
In the following we present the translation strategies for one island of components that are already ordered.
}

In what follows, we describe how a {\em flat} (non-hierarchical), 
{\em connected} diagram is translated. 
If the diagram consists of many disconnected ``islands'', 
we can simply translate each island separately.
Hierarchical diagrams are translated bottom-up: we first translate the 
subsystems, then their parent, and so on.

\ifdefined\techrep 
\subsection{Composition Algorithms for Simulink Components}


In the following we design and implement algorithms that apply the composition operators -- parallel, serial, and feedback -- on any two components $A$ and $B$ for producing the CPT. These algorithms are then used within the translation strategies to obtain the diagram's CPT.

Firstly, we present a decision procedure (Algorithm~\ref{algo:comp}) that determines which composition operators(s) should be applied based on the dependencies between $A$ and $B$. Also, components are ordered such that the number of ``eliminated'' wires is maximized.
If $A$ and $B$ are not connected, parallel composition is applied. Otherwise, serial composition is used, possibly together with feedback if necessary. 

\begin{algorithm}[!h]
\caption{Decision procedure for the composition of two components $A$ and $B$.}\label{algo:comp}
\begin{algorithmic}[1]
\Procedure{compose}{$A, B$}
  \State $A2B\_wires \gets A.out \cap B.in$
  \State $B2A\_wires \gets B.out \cap A.in$
		
  \If{$A2B\_wires = \emptyset$ \textbf{and} $B2A\_wires = \emptyset$}
    \State \textbf{return} \textsc{product}$(A, B)$
  \EndIf
  
  \If{$\vert B2A\_wires \vert \leq \vert A2B\_wires \vert$}
    \If{$\vert B2A\_wires \vert = 0$}
      \State \textbf{return} \textsc{serial}$(A, B)$
    \Else
      \State \textbf{return} \textsc{feedback}(\textsc{serial}$(A, B)$)
    \EndIf
  \Else
    \If{$\vert A2B\_wires \vert = 0$}
      \State \textbf{return} \textsc{serial}$(B, A)$
    \Else
      \State \textbf{return} \textsc{feedback}(\textsc{serial}$(B, A)$)
    \EndIf  
  \EndIf
\EndProcedure
\end{algorithmic}
\end{algorithm}

On the example from Fig.\ref{fig:mex_a}, \textsc{compose}$(A, B)$ will yield the composition of $A$ and $B$ in series encased in a feedback. \textsc{compose}$(B, A)$ gives the feedback of the serial composition of $B$ and $A$, represented in Fig.\ref{fig:mex_b}.\ Fig.\ref{fig:mex_c} will not be produced by the decision procedure.

Each of the algorithms \textsc{parallel}, \textsc{feedback}, and \textsc{serial} creates a new component that replaces within the diagram the components $A$ and $B$. Therefore, we need to compute the list of inputs, outputs and the CPT.


Algorithm~\ref{algo:prod} details the component obtained by applying the parallel composition operator between $A$ and $B$. The lists of inputs/outputs of the new component $res$ are given by the concatenation of the individual lists of inputs and outputs respectively. The CPT is written as $P_A \pop P_B$. On Fig.\ref{fig:mex_c}, \textsc{parallel}$(A, B)$ gives a component with inputs $\{a, b, c\}$, outputs $\{c, d, a\}$, and the previous CPT.

\begin{algorithm}[!h]
\caption{Parallel operator-based composition for two blocks $A$ and $B$.}\label{algo:prod}
\begin{algorithmic}[1]
\Product{$A, B$}
  \State $res.in \gets A.in + B.in$
  
  \State $res.out \gets A.out + B.out$
  

	\State $res.cpt \gets A.cpt \pop B.cpt$
  \State \textbf{return} $res$
\EndProduct
\end{algorithmic}
\end{algorithm}


The \textsc{feedback} algorithm represented in Algorithm~\ref{algo:feedb} applies on a component $A$ with the feedback variables computed as the intersection between its inputs and outputs. The new component has those inputs and outputs which are not feedback variables. The CPT is written as the successive application of the $\fb$ operator on all feedback variables. Recall that the $\fb$ operator applies on one variable at a time. The feedback variables must come first in matching order in the input and output lists. Therefore we need to reorder variables, which is achieved by the first and last expressions of the $\fb$ operator on line~\ref{algo:fb_switch}.

\begin{algorithm}[!h]
\caption{Feedback operator application on a block $A$.}\label{algo:feedb}
\begin{algorithmic}[1]
\Feedback{$A$}
	\State $fdbv \gets A.in \cap A.out$
	
  \State $res.in \gets A.in \setminus fdbv$
  
  \State $res.out \gets A.out \setminus fdbv$
  
  \State $res.cpt \gets [fdbv + res.in \leadsto A.in] \sop A.cpt \sop [A.out \leadsto fdbv + res.out]$ \label{algo:fb_switch}

	\For {i=1 to $\vert fdbv \vert$ }
		\State $res.cpt \gets \fb(res.cpt)$
	\EndFor
	
	\State \textbf{return} $res$
\EndFeedback
\end{algorithmic}
\end{algorithm}

On Fig.\ref{fig:mex_c}, \textsc{feedback}(\textsc{parallel}$(A,B)$) gives the component with input $b$ and output $d$. The explicit CPT is written: $\fb(\fb([a, c, b \leadsto a, b, c] \sop (P_A \pop P_B) \sop [c, d, a \leadsto a, c, d]))$.


The \textsc{serial} function, described by Algorithm~\ref{algo:serial}, proceeds as follows. It computes the unmatched inputs of $B$ and the unmatched outputs of $A$. The inputs of the new component are those of $A$ and the unmatched inputs of $B$. The outputs are those of $B$ and the unmatched outputs of $A$. We transfer the unmatched inputs/outputs with the $\ski$ put in parallel with the according component. The the CPT is given as the serial composition of these expressions, interlaced with a variable reordering expression (line~\ref{algo:serial_switch}) to match the components. 

\begin{algorithm}[!h]
\caption{Serial operator-based composition of two blocks $A$ and $B$.}\label{algo:serial}
\begin{algorithmic}[1]
\Serial{$A, B$}
	\State $B\_unc\_in \gets B.in \setminus A.out$
  \State $res.in \gets A.in + B\_unc\_in$
  
  \State $A\_unc\_out \gets A.out \setminus B.in$
  \State $res.out \gets B.out + A\_unc\_out$
	
	
	\If{$\vert B\_unc\_in \vert = 0$}
			\State $cptA \gets A.cpt$
	\Else
			\State $cptA \gets A.cpt \pop \ski$
	\EndIf

	\If{$\vert A\_unc\_out \vert = 0$}
			\State $cptB \gets B.cpt$
	\Else
			\State $cptB \gets B.cpt \pop \ski$
	\EndIf

  \State $res.cpt \gets cptA \sop [A.out + B\_unc\_in \leadsto B.in + A\_unc\_out] \sop cptB$ \label{algo:serial_switch}
  
  \State \textbf{return} $res$
\EndSerial
\end{algorithmic}
\end{algorithm}

On Fig.\ref{fig:mex_a}, the algorithm creates a component with inputs $\{a, b\}$ and outputs $\{a, d\}$. Since $d$ is an unmatched output of $A$, the CPT of $B$ is composed in parallel with $\ski$. Therefore, we obtain the CPT: $P_A \sop (P_B \pop \ski)$. On Fig.\ref{fig:mex_b}, \textsc{serial}$(B, A)$ yields a component with inputs $\{c, b\}$ and outputs $\{c, d\}$. Now, $b$ is an unmatched input for $A$ and again we use $\ski$ to transfer it. Then, the CPT is: $(P_B \pop \ski) \sop P_A$. To obtain the diagrams CPTs, we apply Algorithm~\ref{algo:feedb}.
\fi 

\ifdefined\techrep
\subsection{Translation Algorithms for HBDs}


Hierarchical block diagrams are translated into CPTs by applying a mixture of the three composition operators (and in consequence algorithms) on the predicate transformers corresponding to the contained components. There are several strategies in which (C)PTs can be assembled together accordingly to the block diagram: (1) by considering all components ``running'' in parallel and reconnecting them through feedback, (2) exploiting the connections between components and maximizing the usage of serial compositions and (3) reorganizing components (after some syntactic sugar constructions) in order to eliminate the feedback operator and use only the parallel and serial ones. As we will show in \S\ref{sec:impl_eval}, a minimal number of applied feedback operators (or even zero) is desirable to optimize the expansion and simplification procedure of CPTs, which is of great interest in performing compatibility and verification checks.
\fi

\subsubsection{Feedback-parallel translation.}
\label{subsubsec:mc}

\ifdefined\techrep
\begin{figure}[!t]
	\centering
	\subfloat[feedback-parallel]{\label{fig:mono_comp}
		\parbox{3cm}{\includegraphics[width=0.15\textwidth]{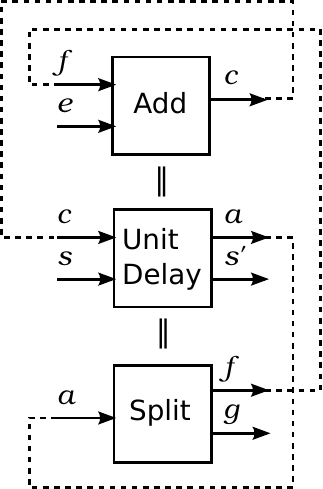}}} \qquad
	\begin{minipage}{0.5\textwidth}
	\centering
	\subfloat[incremental]{\label{fig:inc_comp}
		\includegraphics[width=0.85\textwidth]{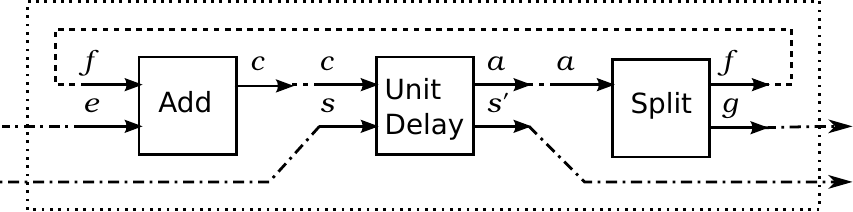}} \\[\baselineskip]
	\subfloat[feedbackless]{\label{fig:nfb_comp}
		\includegraphics[width=0.95\textwidth]{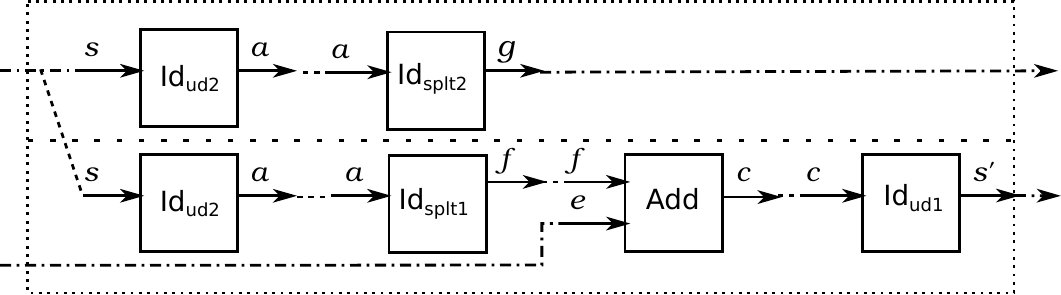}}
	\end{minipage}
	\caption{Translation strategies for the \textsf{DelaySum} subsystem of Fig.\ref{fig:counter_bd}.}\label{fig:comp}
\end{figure}
\else
\begin{figure}[!t]
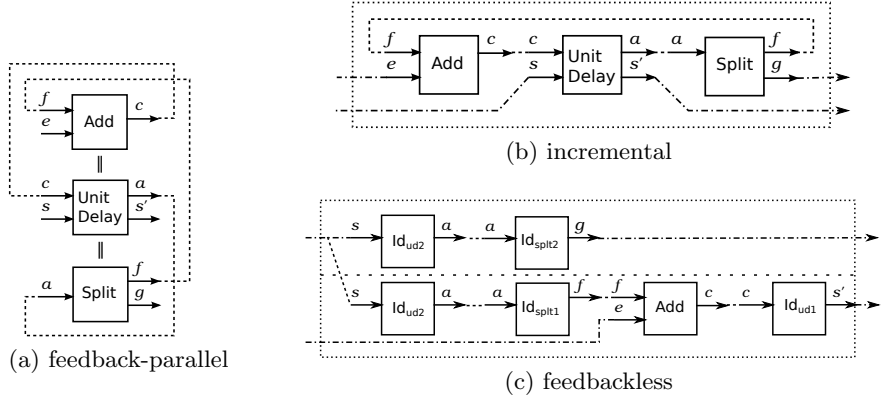

	\centering
	\subfloat[feedback-parallel]{\label{fig:mono_comp}
		\includegraphics[width=0.23\textwidth]{figures/counter_monolithic.pdf}} 
	\hfill
	\begin{minipage}[b]{0.7\textwidth}
	\centering
	\subfloat[incremental]{\label{fig:inc_comp}
		\includegraphics[width=0.85\textwidth]{figures/counter_incremental.pdf}} \\[\baselineskip]
	\subfloat[feedbackless]{\label{fig:nfb_comp}
		\includegraphics[width=0.95\textwidth]{figures/counter_feedbackless.pdf}}
	\end{minipage}
	\caption{Translation strategies for the \textsf{DelaySum} subsystem of Fig.\ref{fig:counter_bd}.}\label{fig:comp}
\end{figure}
\fi

The {\em feedback-parallel translation} strategy (FP) first composes all components
in parallel, 
and then connects outputs to inputs by applying feedback operations 
(with appropriate wiring where necessary). 
%
FP is illustrated in Fig.\ref{fig:mono_comp}, for the \textsf{DelaySum} component of Fig.\ref{fig:counter_bd}. $\mathsf{Split}$ models the junction after wire $a$.

Applying FP on the \textsf{DelaySum} diagram yields the following CPT:
$$
\begin{array}{l}
\mathsf{DelaySum} = \fb^3 ([f,c,a,e,s \leadsto f, e, c, s, a] 
\\[0.6ex]\qquad \sop\ (\mathsf{Add} \pop \mathsf{UnitDelay} \pop \mathsf{Split}) \sop [c, a, s', f, g \leadsto f, c, a, s', g])
\end{array}
$$
where $\fb^3(\cdot)=\fb(\fb(\fb(\cdot)))$ denotes application of the feedback operator 3 times, on the variables $f$, $c$, and $a$, respectively
(recall that $\fb$ works only on one variable at a time, the first input and first output
of the argument transformer). 
In order to apply $\fb^3$ to the parallel composition 
$\mathsf{Add} \pop \mathsf{UnitDelay} \pop \mathsf{Split}$, we first
have to reorder its inputs and outputs, such that the variables on which the feedbacks are applied come first in matching order. 
This is achieved by the {\em rerouting} transformers
$[f,c,a,e,s \leadsto f, e, c, s, a]$ and
$[c, a, s', f, g \leadsto f, c, a, s', g]$.

\subsubsection{Incremental translation.}
\label{subsubsec:ic}

The {\em incremental translation} strategy (IT) composes components one by one,
after having ordered them in topological order according to the dependencies in the diagram.
%
\ifdefined\techrep \else
When composing $A$ with $B$, a decision procedure determines which composition operator(s) should be applied, based on dependencies between $A$ and $B$. If $A$ and $B$ are not connected, parallel composition is applied. Otherwise, serial composition is used, possibly together with feedback if necessary.
\fi

The IT strategy is illustrated in Fig.\ref{fig:inc_comp}. 
First, topological sorting yields the order
$\mathsf{Add},\mathsf{UnitDelay},$ $\mathsf{Split}$.
So IT first composes $\mathsf{Add}$ and $\mathsf{UnitDelay}$.
Since the two are connected with $c$, serial composition is applied,
obtaining the CPT
$$
	\mathsf{ICC1} = (\mathsf{Add} \pop \ski) \sop \mathsf{UnitDelay}
$$
\ifdefined\techrep \else
As in the example in the introduction, $\ski$ is used here to match the number
of outputs of $\mathsf{Add}$ with the number of inputs of $\mathsf{UnitDelay}$. \fi

Next, IT composes $\mathsf{ICC1}$ with $\mathsf{Split}$. This requires both
serial composition and feedback, and yields the final CPT:
$$
	\mathsf{DelaySum} = \fb (\mathsf{ICC1} \sop (\mathsf{Split} \pop \ski))
$$

\begin{figure}[!t]
	\centering
	\includegraphics[scale=0.8]{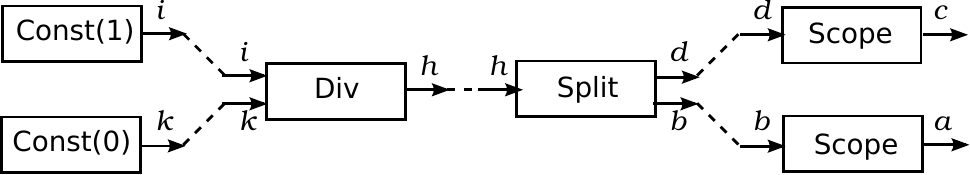} 
	\caption{Diagram \textsf{ConstDiv} illustrating interesting behavior of incremental translation.}\label{fig:hmodel}
\end{figure}

\begin{sloppypar}
It is worth noting that composing systems incrementally in this way might
result in not the most natural compositions. For example,
consider the diagram from Fig.\ref{fig:hmodel}. The ``natural'' CPT for
this diagram is probably: $(\mathsf{Const}(1) \pop \mathsf{Const}(0)) \sop \mathsf{Div} \sop \mathsf{Split} \sop (\mathsf{Scope} \pop \mathsf{Scope})$. 
Instead, IT generates the following CPT: $(\mathsf{Const}(1) \pop \mathsf{Const}(0)) \sop \mathsf{Div} \sop \mathsf{Split} \sop (\mathsf{Scope} \pop \ski) \sop (\ski \pop \mathsf{Scope})$. 
More sophisticated methods may be developed to extract parallelism in the diagram and avoid redundant $\ski$ compositions like in the above CPT. This study is left for future work.
\end{sloppypar}

\subsubsection{Feedbackless translation.}
\label{subsubsec:nfb}

As we shall see in \S\ref{sec:impl_eval}, feedback is problematic for
simplification. The reason is that general feedback 
does not generally preserve the functional predicate transformer
property. It also introduces
formulas with quantifiers, as explained in \S\ref{sec:rcrs}, and quantifier
elimination is not always easy.
We would like therefore to avoid the $\fb$ operator
in the generated CPTs as much as possible. 
The {\em feedbackless translation} strategy (NFBT) avoids $\fb$ altogether.
The key idea is that, since the diagram has no algebraic loops, we should be
able to eliminate feedback and replace it with direct operations on current-
and next-state variables, just like with basic blocks. In particular,
we can {\em decompose} $\mathsf{UnitDelay}$ into two $\ski$ transformers,
denoted $\ski_{\mathsf{ud1}}$ and $\ski_\mathsf{ud2}$:
$\ski_\mathsf{ud1}$ computes the next state from the input,
while $\ski_\mathsf{ud2}$ computes the output from the current state.

Generally, we decompose all components having multiple outputs into several components having each a single output. For each new component we keep only the inputs they depend on, as shown in Fig.\ref{fig:nfb_comp}. Thus, the $\mathsf{Split}$ component from Fig.\ref{fig:inc_comp} is also divided into two $\ski$ components, denoted $\ski_\mathsf{splt1}$ and $\ski_\mathsf{splt2}$.

Decomposing into components with single outputs allows to compute a separate
CPT for each of the outputs. Then we take the parallel composition of these
CPTs to form the CPT of the entire diagram.
\ifdefined\techrep

On the example from Fig.\ref{fig:nfb_comp}, this translation proceeds as follows. It takes the component $\mathsf{UD2}$
and its {\em targets} $\mathsf{Split2}$ and $\mathsf{Split1}$, and replaces them by the new components $\mathsf{UD2} \sop \mathsf{Split2}$ and $\mathsf{UD2} \sop \mathsf{Split1}$. Next it takes the component $\mathsf{UD2} \sop \mathsf{Split1}$
and its target $\mathsf{Add}$, and replaces them by $((\mathsf{UD2} \sop \mathsf{Split1}) \pop \ski) \sop \mathsf{Add}$. In the
final step it takes $((\mathsf{UD2} \sop \mathsf{Split1}) \pop \ski) \sop \mathsf{Add}$ and its target $\mathsf{UD1}$ and 
replaces them by $((\mathsf{UD2} \sop \mathsf{Split1}) \pop \ski) \sop \mathsf{Add} \sop \mathsf{UD1}$. This procedure 
stops when all components have been connected. In general we obtain one component for every output of the systems, and for this example we obtain two
components. In the end these components are composed in parallel to get the CPT.
For this example we obtain:
\else
Doing so on our running example, we obtain:
\fi
$$
\mathsf{DelaySum} = [s, e \leadsto s, s, e] \sop
	\Big(\big(\ski_\mathsf{ud2} \sop \ski_\mathsf{splt2}\big) \pop
	     \big(((\ski_\mathsf{ud2}\sop \ski_\mathsf{splt1}) \pop \ski) \sop \mathsf{Add} \sop \ski_\mathsf{ud1}\big)\Big)
$$ 
Because $\ski_\mathsf{ud1}$, $\ski_\mathsf{ud2}$, $\ski_\mathsf{splt1}$ and
$\ski_\mathsf{splt2}$ are all $\ski$s, and $\ski\sop A = A\sop\ski= A$ and 
$\ski \pop \ski = \ski$ (thanks to polymorphism), this CPT is reduced to $\mathsf{DelaySum} = [s, e \leadsto s, s, e] \sop (\ski \pop \mathsf{Add})$. The direct generation of this simpler CPT is possible, and part of future work.

%

%% file: evaluation.tex

\section{Implementation and Evaluation}
\label{sec:impl_eval}

We implemented the translation algorithms presented in \S\ref{sec:transformation}
into a prototype tool called the \textsc{simulink2isabelle} compiler, downloadable from \url{http://users.ics.aalto.fi/iulia/sim2isa.shtml}.
In this section we present the tool and report evaluation results 
on several case studies, including an industrial-grade benchmark by Toyota~\cite{jin2014ARCH,JinDKUB14}.

\subsection{Toolset}
\label{subsec:tool}

The \textsc{simulink2isabelle} compiler, written in Python, takes as input Simulink files in XML
format and produces valid Isabelle theories that can be subjected to compatibility checking and verification. 
The compiler currently handles a sufficiently large subset of the Simulink libraries able to express industrial-grade models such as the Toyota benchmarks. Extending the compiler to handle even more basic blocks is ongoing work.

During the parsing and preprocessing phase of the input Simulink file, the tool performs a set of checks, including algebraic loop detection, 
unsupported blocks and/or block parameters, malformed blocks (e.g., a function
block referring to a nonexistent input), etc., and issues possible warnings/errors.

The tool implements all three translation strategies, and takes two additional options: \texttt{flat} (flatten diagram) and \texttt{io} (intermediate outputs).
Option \texttt{flat} flattens the hierarchy of the HBD and produces a single
diagram consisting only of basic blocks (no subsystems), on which the translation is then applied. \ifdefined\techrep Note that FP and IT are by design compositional, i.e., they preserve the hierarchical structure of the model. However, using the \texttt{flat} option discards this feature. \fi
Option \texttt{io} generates and names all intermediate CPTs produced during
the translation process. These names are then used in the CPT for the top-level
system, to make it shorter and more readable.
In addition, the intermediate CPTs can be expanded and simplified incrementally
by Isabelle, and used in their simplified form when computing the CPT for the
next level up. This generally results in more efficient simplification.
Another benefit of producing intermediate CPTs is the detection of 
incompatibilities early during the simplification phase. Moreover, this 
indicates the group of components at fault and helps localize the error.


Isabelle is a proof assistant, but not a fully-automatic tool.
We have therefore implemented in Isabelle a set of functions (keyword
\textbf{simulink})
which result in the generated CPTs being expanded and simplified automatically.
\ifdefined\techrep
By {\em expansion} we mean replacing the serial, parallel, and feedback
composition operators based on the Lemmas~\ref{lem:serial-prod} and~ \ref{lem:feedback}.
\else
By {\em expansion} we mean replacing the serial, parallel, and feedback
composition operators by their definitions (\ref{eq:serial}),
(\ref{eq:parallel}), (\ref{eq:feedback}).
\fi
By {\em simplification} we mean simplifying the resulting formulas, e.g.,
by eliminating quantifiers and internal variables. The goal is to obtain
for the top-level system a single atomic predicate transformer which only
refers to the external input, output, and state/next state variables of 
the system (and not to the internal wires of the diagram).

For instance, when executed on our running example (Fig.\ref{fig:counter}) with the IT option, the tool produces the Isabelle code:
$$
    \begin{array}{l}
    \textbf{simulink } \mathsf{DelaySum} =
      \fb((\mathsf{Add}\pop\ski)\sop \mathsf{UnitDelay}\sop(\mathsf{Split} \pop \ski))\\[0.7ex]
	\textbf{simulink } \mathsf{Counter} = (\mathsf{Const}(1) \pop \ski) \sop \mathsf{DelaySum} \sop (\mathsf{Scope} \pop \ski)
	\end{array}
$$
When executed in Isabelle, this code automatically generates the definitions 
(\ref{eq:counter:def}) as well as the simplification theorems 
(\ref{eq:counter:simp}), and automatically proves these theorems.

As another example, when we run the tool on the example of Fig.\ref{fig:hmodel},
we obtain the theorem $\mathsf{ConstDiv} = \{x:\mathsf{false}\}$, which states
that the system has no legal inputs. This indicates incompatibility, due to
performing a division by zero.

\ifdefined\techrep
The generated $\textbf{simulink}$ declarations preserve the name of the block instance they map, therefore enforcing the traceability of the translation.
\else
For more details about the Isabelle code, the reader is referred to \cite{report}.
\fi

\subsection{Evaluation} 
\label{subsec:evaluation}

We present evaluation results from two case studies: the running example 
(Fig.\ref{fig:counter}) and the Fuel Control System (FCS) model described in \cite{jin2014ARCH,JinDKUB14}. FCS solves the problem of maintaining the ratio of air mass and injected fuel at the stoichiometric value \cite{CookSBKPG06}, i.e., enough air is provided to completely burn the fuel in a car's engine. This control problem has important implications on lowering pollution and improving engine performance.
Three designs are presented in~\cite{jin2014ARCH,JinDKUB14}, modeled in Simulink, Hybrid I/O Automata~\cite{LynchSV03} and Polynomial Hybrid I/O Automata~\cite{FrehseHK04}. We evaluate our approach on the Simulink model, available
from \url{http://cps-vo.org/group/ARCH/benchmarks}.
The model has a 3-level hierarchy with a total of 70 block instances (65 basic blocks and 5 subsystems), connected by 82 wires, of which 8 feedbacks.

We evaluate the three translation strategies with the options:
FP without/with flattening (PHBD/FHBD), 
IT without/with flattening and without/with \texttt{io} option (IO), 
and NFBT. 
NFBT by construction generates intermediate outputs and 
does not preserve the structure of the hierarchy in the result, thus, its
result is identical with/without the options.
The results from the running example are shown in Table~\ref{table:eval_counter} and from the FCS model in Table~\ref{table:eval_fcs}. 
Table~\ref{table:eval_subsys} shows results from the largest flat subsystem 
of the FCS model. This subsystem has 35 basic blocks and 47 wires.

 The evaluation criteria are:
(1) $\mathcal{L}_{\mbox{cpt}}$: length of the produced CPTs (\# characters),
(2) $\mathcal{N}_{\mbox{cpt}}$: number of generated CPTs,
(3) readability,
(4) $\mathcal{T}_{\mbox{exp}}$ and $\mathcal{P}_{\mbox{exp}}$: times required for the expansion of CPTs and for the printing of the top-level formula,
(5) $\mathcal{N}_{\mbox{quant}}$: number of quantifiers in the top-level formula,
(6) $\mathcal{T}_{\mbox{tot}}$: total time needed for expansion 
and simplification,
(7) $\mathcal{P}_{\mbox{simp}}$: time to print the simplified formula,
(8) $\mathcal{L}_{\mbox{simp}}$: length of the simplified formula (\# chars).
All times are in seconds.
The translation time itself (i.e., the time to process the Simulink file
and generate the Isabelle file) is negligible for all three strategies (at most one fifth of a second) and therefore not reported.
We present separately times to expand/simplify and times to print the formulas,
since printing takes significant time in the Isabelle/ML framework.

\begin{table*}[!t]
\centering
\begin{tabular}{ c | c | c | c | c | c | c | c | }
\cline{2-8}
 & \multicolumn{2}{ c |}{FP} & \multicolumn{4}{ c |}{IT} & \multirow{2}{*}[5.5pt]{~NFBT~} \\ \cline{2-7} 
 & PHBD & FHBD & PHBD & FHBD & ~IO-PHBD~ & ~IO-FHBD~ &  \\ \hline
\multicolumn{1}{| c |}{$\mathcal{L}_{\mbox{cpt}}$} & 731 & 598 & 1317 & 1546 & 1325 & 1303 & 2455 \\ \hline
\multicolumn{1}{| c |}{$\mathcal{N}_{\mbox{cpt}}$} & 2 & 1 & 2 & 1 & 6 & 6 & 4 \\ \hline
\multicolumn{1}{| c |}{~Readability~} & ~medium~ & ~medium~ & ~medium~ & ~medium~ & high & high & low \\ \hline
\multicolumn{1}{| c |}{$\mathcal{T}_{\mbox{exp}}$} & 0.095 & 0.114 & 0.120 & 0.154 & 0.193 & 0.199 &0.102  \\ \hline
\multicolumn{1}{| c |}{$\mathcal{P}_{\mbox{exp}}$} & 0.013 & 0.016 & 0.024 & 0.017 & 0.018 & 0.017 & 0.107 \\ \hline
\multicolumn{1}{| c |}{$\mathcal{N}_{\mbox{quant}}$} & 11 & 9 & 19 & 21 & 20 & 20 & 0 \\ \hline
\multicolumn{1}{| c |}{$\mathcal{T}_{\mbox{tot}}$} & 0.247 & 0.312 & 0.220 & 0.338 & 0.263 & 0.274 & 0.072 \\ \hline
\multicolumn{1}{| c |}{$\mathcal{P}_{\mbox{simp}}$} & 0.01 & 0.0 & 0.0 & 0.0 & 0.0 & 0.0 & 0  \\ \hline
\end{tabular}
\caption{Evaluation results for the running example (Fig.\ref{fig:counter}).}
\label{table:eval_counter}
\end{table*} 
\ifdefined\techrep \else \vspace*{-0.2cm} \fi

\begin{table*}[!t]
\centering
\begin{tabular}{ c | c | c | c | c | c | c | c | }
\cline{2-8}
 & \multicolumn{2}{ c |}{FP} & \multicolumn{4}{ c |}{IT} & \multirow{2}{*}[5.5pt]{NFBT} \\ \cline{2-7} 
 & ~PHBD~ & ~FHBD~ & ~PHBD~ & ~FHBD~ & ~IO-PHBD~ & ~IO-FHBD~ &  \\ \hline
\multicolumn{1}{| c |}{$\mathcal{L}_{\mbox{cpt}}$} & 17191 & 15955 & 141989 & 167291 & 155871 & 189581 & 33372\\ \hline
\multicolumn{1}{| c |}{$\mathcal{N}_{\mbox{cpt}}$} & 6 & 1 & 6  & 1 & 84 & 84 & 101 \\ \hline
\multicolumn{1}{| c |}{$\mathcal{T}_{\mbox{exp}}$} & $\infty$ & $\infty$ & $\infty$ & $\infty$ & 1809.184 & $\infty$ & 5.564 \\ \hline
\multicolumn{1}{| c |}{$\mathcal{P}_{\mbox{exp}}$} & - & - & - & - & $\infty$ & - & 27.017 \\ \hline
\multicolumn{1}{| c |}{~$\mathcal{N}_{\mbox{quant}}$~} & - & - & - & - & - & - & 0 \\ \hline
\multicolumn{1}{| c |}{$\mathcal{T}_{\mbox{tot}}$} & - & - & - & - & 3393.787 & 3892.872 & 14.089\\ \hline
\multicolumn{1}{| c |}{$\mathcal{P}_{\mbox{simp}}$} & - & - & - & - & 1.756 & 2.795 & 1.140  \\ \hline
\multicolumn{1}{| c |}{$\mathcal{L}_{\mbox{simp}}$} & - & - & - & - & 41047 & 45761 & 21243 \\ \hline
\end{tabular}
\caption{Evaluation results for the FCS Simulink model~\cite{jin2014ARCH,JinDKUB14}.}
\label{table:eval_fcs}
\end{table*} 

\begin{table*}[!t]
\centering
\begin{tabular}{ c | c | c | c | c | }
\cline{2-5}
 & \multirow{2}{*}[5.5pt]{~FP~} & \multicolumn{2}{ c |}{IT} & \multirow{2}{*}[5.5pt]{~NFBT~} \\ \cline{3-4} 
 & & ~PHBD~ & ~IO-PHBD~ &  \\ \hline
\multicolumn{1}{| c |}{$\mathcal{L}_{\mbox{cpt}}$} & 13042 & 88059 & 85116 & 19258 \\ \hline
\multicolumn{1}{| c |}{$\mathcal{N}_{\mbox{cpt}}$} & 1 & 1 & 47  & 56 \\ \hline
\multicolumn{1}{| c |}{$\mathcal{T}_{\mbox{exp}}$} & 32.197 & 287.874  & 243.86  & 3.029 \\ \hline
\multicolumn{1}{| c |}{$\mathcal{P}_{\mbox{exp}}$} & 456.758 & 371.601  & 276.799  & 4.527 \\ \hline
\multicolumn{1}{| c |}{$\mathcal{L}_{\mbox{exp}}$} & 176229 & 176229 & 216674 & 209125 \\ \hline
\multicolumn{1}{| c |}{~$\mathcal{N}_{\mbox{quant}}$~} & 176 & 176 & 193 & 0 \\ \hline
\multicolumn{1}{| c |}{$\mathcal{T}_{\mbox{tot}}$} & 1689.427 & 1401.053 & 969.684 & 5.524 \\ \hline
\multicolumn{1}{| c |}{$\mathcal{P}_{\mbox{simp}}$} & 0.603 & 0.571 & 0.525 & 0.327 \\ \hline
\multicolumn{1}{| c |}{$\mathcal{L}_{\mbox{simp}}$} & 35115 & 17577 & 17577 & 17303 \\ \hline
\end{tabular}
\caption{Evaluation results for the largest subsystem of the FCS model.}
\label{table:eval_subsys}
\end{table*} 
\ifdefined\techrep \else \vspace*{-0.2cm} \fi

Readability is subjective, of course, and our measures are relative. 
Readability is higher in the incremental strategy with \texttt{io} option,
since the intermediate outputs allows to parse the result step by step.
Readability is low in NFBT because this method decomposes blocks and
does not preserve the hierarchy of the original model.
For this reason, and even though NFBT is superior in terms of efficiency,
the other methods are still interesting to pursue and improve as part of
future work.

The $\infty$ symbol denotes timeout after two hours of computation.
The reason Isabelle fails to expand or print some formulas is that they
become too large. 
Also, expansion involves more than just inlining, e.g.,
renaming variables appropriately due to quantifiers.
We note that
$\mathsf{feedback}(\{p\} \circ [r])$ expands to a formula with length roughly equal to $(3\cdot\vert p \vert+\vert r \vert) + (\vert p \vert + 2\cdot\vert r \vert)$. 
Similarly, $\fb$ introduces $3 + 3 \cdot \mathcal{N}_{\mbox{quant}}(p)+ \mathcal{N}_{\mbox{quant}}(r)$ quantifiers in the precondition and $3 + \mathcal{N}_{\mbox{quant}}(p) + 2 \cdot \mathcal{N}_{\mbox{quant}}(r)$
in the relation.
Note also that a feedback wire can transfer an array of $n$ values, which is translated by our tool as $n$ successive applications of $\fb$.

Observe (Table~\ref{table:eval_fcs})
that with IO-PHBD Isabelle manages to expand the CPTs (into an internal
data structure) after about 30 mins, but not to print the expanded formula.
However, the expanded formula can be simplified in less than 30 mins and the
simplified formula can be printed in less than 2 secs.
In the case of NFBT, everything (expansion+simplification+printing) takes
less than a sec.
The final simplified formulas are in all cases quantifier-free.

%% file: rwork.tex
\section{Related Work}
\label{sec:rwork}

A plethora of work exists on translating Simulink models to various 
target languages, for verification purposes or for code-generation purposes.
Although some of these works have ultimate goals similar to ours
(e.g., verification), to the best of our knowledge, none of them
studies the same problem that we study in this paper: the different ways to
translate HBDs into a compositional algebra with
serial, parallel, and feedback composition operators.

Primarily focusing on verification and targeting discrete-time fragments of
Simulink, existing works describe translations to BIP~\cite{SfyrlaTSBS10}, NuSMV~\cite{MeenakshiBR06}, or Lustre~\cite{TripakisSCC05}. 
Other works study transformation of continuous-time Simulink to Timed Interval Calculus~\cite{ChenDS09}, Function Blocks~\cite{YangV12}, I/O Extended Finite Automata \cite{ZhouK12}, or Hybrid CSP~\cite{ZouZWFQ13}. 
The Stateflow module of Simulink, which allows to model hierarchical state machines, has been the subject of translation to hybrid automata~\cite{AgrawalSK04,ManamcheriMBC11}. 

Contract-based frameworks for Simulink are described in~\cite{Bostrom11,RoyS11}.
Although these works are related to ours, they do not solve the
same problem as explained above.
\cite{Bostrom11} uses pre/post-conditions as contracts for discrete-time 
Simulink blocks and SDF graphs~\cite{LeeMesserschmitt87} to represent
Simulink diagrams.
%
Then sequential code is generated from the SDF graph, and the code
is verified using traditional refinement-based techniques~\cite{back:wright:98}
and the Z3 SMT solver~\cite{DeMouraB08}.
%
In~\cite{RoyS11} Simulink blocks are annotated with rich types
(separate constraints on inputs and outputs, but no relations between
inputs and outputs which is possible in our framework).
Then the SimCheck tool extracts verification conditions from the Simulink model and the annotations, and submits them to the Yices SMT solver~\cite{SRI:yices:tool} for verification.

Modular code generation methods for Simulink models are described
in~\cite{LublinermanTripakisDATE08,LublinermanTripakisPOPL09}.
The main technical problem solved there is how to cluster subsystems
in as few clusters as possible without introducing false input-output
dependencies.

%% file: conclusion.tex
\section{Conclusion}
\label{sec:conclusion}

In this paper we tackle the problem of translating Simulink diagrams (and
more generally, hierarchical block diagrams) into a compositional framework:
an algebra of atomic systems composed in series, parallel, or feedback.
The challenge comes from the fact that there are many ways to translate
a diagram into a term in this algebra, with different tradeoffs.
Three translation strategies are proposed and implemented in a compiler 
that generates verifiable Isabelle code. These strategies are illustrated 
on a simple example and evaluated on a real-life Simulink model from Toyota.

\ifdefined\techrep
Two intuitive translation strategies are first considered: the feedback-parallel translation and the incremental translation. These strategies have the benefit of producing relatively readable composite predicate transformers, which can
be traced back to the structure of the original diagram.
 Unfortunately, expansion and simplification of the CPTs resulting from these translations do not scale well in Isabelle.
The culprits are the length of the predicate transformers and the feedback composition operators that introduce quantifiers and are resource consuming. A third strategy -- feedbackless -- was designed to handle this scalability issue by avoiding feedback altogether. The measurements obtained show that the feedbackless method is by far the most efficient on all examples we tried. The drawback 
of the feedbackless method is that it decomposes the blocks of the original diagram, which renders this blocks difficult to identify in the produced CPTs. In consequence, the resulting expressions are harder to understand and diagnose in case of errors. 
\fi

Future work directions include:
(1) studying other translation strategies;
(2) improving the automated simplification methods within Isabelle or other
solvers;
(3) extending the tool to handle a larger subset of Simulink;
(4) extending the tool with automatic verification methods (model-checking
requirements against the top-level CPT);
and 
(5) extending the tool with fault localization methods whenever the
compatibility or verification checks fail.

Another interesting question regards the semantical equivalence of the CPTs
generated by the various strategies and options. 
We proved in Isabelle that the final simplified formulas are equivalent,
for the running example as well as for several other examples and subsystems of the FCS model, including the largest subsystem reported in Table~\ref{table:eval_subsys}. It is future work to prove a meta-theorem that states that these equivalences hold by construction on any algebraic-loop-free example.